\documentclass[prd,amsmath,amssymb,longbibliography,superscriptaddress,twocolumn,nofootinbib,10pt]{revtex4}

\pdfoutput=1

\usepackage{graphicx}
\usepackage{dcolumn}
\usepackage{bm}
\usepackage{amssymb}
\usepackage{latexsym}
\usepackage{booktabs}
\usepackage{amsmath}
\usepackage{multirow}
\usepackage{xcolor}
\usepackage[utf8]{inputenc}

\newcommand{\LCDM}{\Lambda\text{CDM}}
\newcommand{\Mpc}{\mathrm{~km~s^{-1}~Mpc^{-1}}}
\newcommand{\clight}{c_{\rm light}}
\newcommand{\be}{\begin{equation}}
\newcommand{\ee}{\end{equation}}

\usepackage[colorlinks=true, linkcolor=blue, citecolor=blue]{hyperref}

\begin{document}

\title{Late-time cosmological constraints on three holographic dark energy models with DESI DR2 BAO and Type Ia supernovae}
\author{Jiayuan Huang}
\affiliation{School of Physics and Optoelectronic, Yangtze University, Jingzhou 434023, China;}
\author{Tonghua Liu}
\email{liutongh@yangtzeu.edu.cn}
\affiliation{School of Physics and Optoelectronic, Yangtze University, Jingzhou 434023, China;}
\author{Chenggang Shao}
\email{cgshao@hust.edu.cn}
\affiliation{School of Physics and Optoelectronic, Yangtze University, Jingzhou 434023, China;}

\begin{abstract}
We constrain three holographic-inspired dark energy models, namely holographic dark energy (HDE), agegraphic dark energy (ADE), and Ricci dark energy (RDE), using late-time observations from cosmic chronometers, Type Ia supernovae (SNe Ia), DESI DR2 baryon acoustic oscillations (BAO), and  {redshift-space distortion (RSD) growth measurements}. Five data combinations are considered: $H(z)+$Pantheon+, $H(z)+$DESI DR2+Pantheon+, $H(z)+$DESI DR2+DES-Dovekie, $H(z)+$DESI DR2+DESY5, and  {$H(z)+$DESI DR2+DES-Dovekie+RSD}. We perform Bayesian Markov chain Monte Carlo parameter estimation and compare the models with AIC and BIC. In the BAO-included combinations, HDE gives $H_0\simeq67.3$--$68.0~\Mpc$, $\Omega_{m0}\simeq0.270$--$0.272$, and $c\simeq1$, indicating an expansion history close to the de Sitter boundary rather than a robust phantom regime. ADE yields a stable agegraphic parameter $n\simeq2.78$--$2.81$, while RDE gives $\gamma\simeq0.53$--$0.55$ and persistently favors a low matter density, $\Omega_{m0}\simeq0.215$--$0.219$.  {Treating $r_d$ as a free parameter reveals a strong negative correlation between $H_0$ and $r_d$, and the RSD-included combination provides a growth-level consistency check through $f\sigma_8(z)$ without constituting a full perturbative stability analysis.} None of the three models significantly alleviates the Hubble tension. Overall, HDE shows the most balanced phenomenological behavior among the three models, although current late-time data do not decisively prefer it over $\LCDM$.
\end{abstract}
\date{\today}
\maketitle

\section{Introduction}
\label{sec1}
The discovery of cosmic acceleration through Type Ia supernovae (SNe Ia) established the existence of dark energy and has become one of the central open questions in modern cosmology \cite{Riess1998,Perlmutter1999}. Within general relativity, the cosmological constant $\Lambda$ provides the simplest explanation for late-time acceleration and forms the basis of the standard $\LCDM$ cosmological model \cite{Lahav2020}. This model has achieved remarkable success in describing cosmic microwave background (CMB) anisotropies, baryon acoustic oscillations (BAO), large-scale structure clustering, and SNe Ia distances. Nevertheless, it still faces long-standing theoretical challenges, including the cosmological constant problem \cite{Weinberg1989} and the coincidence problem \cite{Steinhardt1997}.

In parallel with these theoretical issues, increasingly precise observations have revealed possible tensions within the $\LCDM$ framework. The most prominent example is the Hubble tension \cite{Di2021}, namely the discrepancy between the value of $H_0$ inferred from early-universe CMB observations under the $\LCDM$ assumption and that measured by local distance-ladder methods. The Planck Collaboration reports $H_0=67.4\pm0.5~\Mpc$ \cite{2020A&A...641A...6P}, whereas the SH0ES Collaboration obtains a higher local value, $H_0=73.04\pm1.04~\Mpc$ \cite{2022ApJ...934L...7R}. Other independent or partially independent determinations give values spanning a broad range, including the Megamaser Cosmology Project result $H_0=73.9\pm3.0~\Mpc$ \cite{2020ApJ...891L...1P}, the TDCOSMO strong-lensing time-delay constraint $H_0=67.4^{+4.1}_{-3.2}~\Mpc$ \cite{2020A&A...643A.165B}, and the tip of the red giant branch calibration $H_0=69.6\pm2.5~\Mpc$ \cite{2020ApJ...891...57F}. These measurements do not point to a single simple resolution; instead, they suggest that the tension may involve residual observational systematics, calibration differences, or limitations of the standard cosmological model. For recent reviews and discussions of the Hubble tension, see Refs.~\cite{2021APh...13102605D,2022JHEAp..34...49A,2022NewAR..9501659P,2019PhRvL.122v1301P,2019PhRvL.122f1105F,2021ApJ...912..150D,2020ApJ...895L..29L,2025PhRvD.112l3539L,2022ApJ...939...37L,2023ApJS..264...46L} and references therein. This situation has motivated extensive studies of models beyond $\LCDM$, including early dark energy, interacting dark sectors, modified gravity, and dynamical dark energy scenarios \cite{2019NatAs...3..891V,2021CQGra..38o3001D,2023Univ....9..393V,2020PhRvL.125r1302J,2020MNRAS.496L..91E,2021ApJ...919...16F}.

Recent BAO measurements have further sharpened the observational test of late-time cosmic expansion. The Dark Energy Spectroscopic Instrument (DESI) has provided high-precision spectroscopic measurements of galaxies, quasars, and Lyman-$\alpha$ forest tracers over a broad redshift range. Its first data release (DESI DR1) already delivered percent-level BAO distance constraints \citep{2025JCAP...02..021A}, while the second data release (DESI DR2) further improves the statistical power through enlarged sky coverage and refined redshift measurements \cite{Abdul2025}. In combination with CMB and SNe Ia data, recent DESI analyses have reported indications that the dark energy equation of state may deviate from a cosmological constant, often described phenomenologically by the Chevallier--Polarski--Linder (CPL) parametrization \cite{Chevallier:2000qy,Linder:2002et}. Although such hints remain sensitive to data choices and model assumptions, they have renewed interest in physically motivated dynamical dark energy models.  This has prompted researchers to investigate more theoretically grounded explanations for the observed cosmic acceleration. For a nonexhaustive set of references on this topic, see the following \cite{Giare:2024gpk,Giare:2024smz,Giare:2024oil,Li:2025ula,Li:2024qso,Jiang:2024viw,Li:2025owk,2025arXiv250321652S,Li:2026xaz,Li:2025ops,Fazzari:2025lzd,Ozulker:2025ehg,Wolf:2025jed,Wolf:2025acj,Du:2025xes,Du:2026cly,Teixeira:2025czm,vanderWesthuizen:2025rip,Cheng:2025lod,Silva:2025hxw,Pedrotti:2025ccw,Wang:2026kbg,Zhou:2025nkb,Zhang:2025dwu,RoyChoudhury:2025iis,Luciano:2025elo,Paliathanasis:2026ymi,Cheng:2025yue,DESI:2025fii,Keeley:2025stf,DESI:2025wyn,2026EPJC...86..391W,2025EPJC...85.1351L,2025ApJ...988..243L,2025ApJ...982...71Y,2026ApJ..1001L..21L}.

The CPL parametrization is useful for model-independent phenomenology, but it does not by itself specify the microscopic or gravitational origin of dark energy. This motivates complementary studies of theoretically grounded models, in which the dark energy density is tied to a physical length scale or curvature invariant. Among these possibilities, holographic dark energy scenarios are particularly attractive because they are inspired by the holographic principle and quantum-gravity-related entropy bounds \cite{1993gr.qc....10026T,1995JMP....36.6377S}. Applied to effective field theory, the holographic argument introduces a relation between ultraviolet and infrared cutoffs, leading naturally to a dynamical vacuum energy component.

The original holographic dark energy (HDE) model was proposed by  \citet{2004PhLB..603....1L}, who identified the infrared cutoff with the cosmic future event horizon. This model has been widely studied and can provide a phenomenologically viable description of late-time cosmic acceleration \cite{2004JCAP...08..006H,2005PhRvD..72d3524Z,2014SCPMA..57..387Z,2010PhR...493....1C,2017PhR...696....1W}. Related holographic-inspired constructions include agegraphic dark energy (ADE), in which the infrared scale is associated with the conformal age of the Universe \cite{2008PhLB..660..113W}, and Ricci dark energy (RDE), in which the dark energy density is linked to the spacetime Ricci curvature \cite{2015JMPh....6..327S}. These three models share a common holographic motivation but predict different late-time expansion histories and different evolutions of the dark energy equation of state.

In this work, we perform a comparative late-time constraint analysis of HDE, ADE, and RDE using cosmic chronometer $H(z)$ measurements, SNe Ia samples, DESI DR2 BAO data,  {and an extended RSD growth compilation}. We consider five data combinations: $H(z)+$Pantheon+, $H(z)+$DESI DR2+Pantheon+, $H(z)+$DESI DR2+DES-Dovekie, $H(z)+$DESI DR2+DESY5, and  {$H(z)+$DESI DR2+DES-Dovekie+RSD}. The sound horizon $r_d$ is treated as a free parameter in BAO-included combinations, allowing a late-time analysis without imposing a CMB-calibrated sound-horizon prior.  {The RSD combination is included as a growth-level consistency test through $f\sigma_8(z)$, with $\sigma_{8,0}$ treated as an additional nuisance parameter.} We derive posterior constraints using Bayesian Markov chain Monte Carlo sampling and compare the models with the Akaike information criterion and Bayesian information criterion. Our goal is to assess whether any of these holographic-inspired dark energy models can provide a statistically competitive and physically consistent alternative to the standard $\LCDM$ cosmology in light of current late-time and growth observations.

\section{Theoretical Models}
\label{sec:models}
In this section, we introduce three typical holographic-inspired dark energy models adopted in our analysis. We assume a spatially flat Friedmann--Robertson--Walker (FRW) cosmological background filled with cold dark matter, radiation, and dark energy. We define the dimensionless Hubble parameter as
\begin{equation}
E(z) \equiv \frac{H(z)}{H_0},
\end{equation}
which characterizes the cosmic expansion history. For all models, $\Omega_{\rm de}(z)$ denotes the fractional dark energy density at redshift $z$, and the present radiation density parameter $\Omega_{r0}$ is fixed to the standard CMB value. The difference among these models originates from the physical construction of dark energy, which determines the form of $E(z)$ through the Friedmann equation.

\subsection{Holographic Dark Energy Model}
The HDE model arises from combining the holographic principle with effective quantum field theory \cite{Cohen1999}. The core physical idea is that the vacuum energy within a characteristic cosmic scale cannot exceed the mass of a black hole with the same size, which gives an upper bound on the ultraviolet energy cutoff. Saturating this bound, the dark energy density is parametrized as
\begin{equation}
\rho_{\rm de}=3c^{2}M_{\rm p}^{2}L^{-2},
\end{equation}
where $c$ is a dimensionless model parameter, $M_{\rm p}=1/\sqrt{8\pi G}$ is the reduced Planck mass, and $L$ serves as the infrared cutoff scale.

In the standard HDE framework, the future event horizon is chosen as the infrared cutoff $L$ \cite{2004PhLB..603....1L}. For a flat FRW universe, the Friedmann equation constrains the present dark energy density parameter via spatial flatness: $\Omega_{\rm de0}=1-\Omega_{m0}-\Omega_{r0}$. Once the redshift evolution of $\Omega_{\rm de}(z)$ is determined, the dimensionless Hubble parameter takes the form
\begin{equation}
E_{\rm HDE}^{2}(z)
=
\frac{\Omega_{m0}(1+z)^3+\Omega_{r0}(1+z)^4}{1-\Omega_{\rm de}(z)}.
\label{eq:E_HDE}
\end{equation}

The dynamical evolution of $\Omega_{\rm de}(z)$ is governed by
\begin{equation}
\frac{{\rm d}\Omega_{\rm de}}{{\rm d}z}
=
-\frac{\Omega_{\rm de}}{1+z}
\left[
(1-\Omega_{\rm de})
\left(
1+\frac{2\sqrt{\Omega_{\rm de}}}{c}
\right)
+\Omega_r(z)
\right],
\label{eq:HDE_evolution}
\end{equation}
where $\Omega_r(z)$ is the fractional radiation density at redshift $z$. The equation-of-state parameter of HDE reads
\begin{equation}
w_{\rm de}(z)
=
-\frac{1}{3}
-
\frac{2\sqrt{\Omega_{\rm de}(z)}}{3c}.
\label{eq:w_HDE}
\end{equation}
The value of $c$ directly determines the nature of dark energy: $c>1$ corresponds to quintessence-like behavior in the dark-energy-dominated limit, $c=1$ approaches the de Sitter limit, while $c<1$ leads to phantom evolution. Radiation contributions become negligible at late times, but they are retained in the numerical evolution for consistency.

\subsection{Agegraphic Dark Energy Model}
The ADE model is constructed based on spacetime quantum fluctuations and the K\'arolyh\'azy uncertainty relation \cite{2008PhLB..660..113W}. Unlike HDE, which adopts the future event horizon as the cutoff, ADE relates the dark energy density to the conformal age of the universe, avoiding explicit dependence on future cosmic evolution and providing a causal realization of holographic-inspired dark energy.

The dark energy density of ADE is given by
\begin{equation}
\rho_{\rm de}
=
3n^{2}M_{\rm p}^{2}\eta^{-2},
\end{equation}
where $n$ is a dimensionless model parameter and $\eta$ denotes the conformal time of the universe,
\begin{equation}
\eta
=
\int_0^a \frac{{\rm d}a'}{H a'^2}.
\end{equation}
In this work, the ADE model refers to the new agegraphic dark energy scenario based on the conformal age of the Universe.
Similar to HDE, the dimensionless Hubble parameter can be expressed as
\begin{equation}
E_{\rm ADE}^{2}(z)
=
\frac{\Omega_{m0}(1+z)^3+\Omega_{r0}(1+z)^4}{1-\Omega_{\rm de}(z)}.
\label{eq:E_ADE}
\end{equation}

The redshift evolution of $\Omega_{\rm de}(z)$ is solved numerically. Including the radiation correction, it can be written as
\begin{equation}
\frac{{\rm d}\Omega_{\rm de}}{{\rm d}z}
=
\frac{\Omega_{\rm de}(1-\Omega_{\rm de})}{1+z}
\left[\frac{2(1+z)\sqrt{\Omega_{\rm de}}}{n}
-3-F(z)
\right],
\label{eq:ADE_evolution}
\end{equation}
where
\begin{equation}
F(z)
=
\frac{\Omega_{r0}(1+z)}
{\Omega_{m0}+\Omega_{r0}(1+z)} .
\end{equation}
The equation-of-state parameter is
\begin{equation}
w_{\rm de}(z)
=
-1
+
\frac{2(1+z)\sqrt{\Omega_{\rm de}(z)}}{3n}.
\label{eq:w_ADE}
\end{equation}
In our numerical analysis, we do not treat $\Omega_{m0}$ as an independent free parameter for ADE. Instead, it is fixed by the spatial flatness condition and the normalization $E(0)=1$ for given values of $H_0$ and $n$.

\subsection{Ricci Dark Energy Model}
The RDE model identifies the infrared cutoff scale with the spacetime Ricci scalar curvature in the FRW background \cite{2015JMPh....6..327S}. The Ricci scalar is jointly determined by the Hubble parameter and its time derivative, linking dark energy to the background spacetime geometry. The RDE energy density is written as
\begin{equation}
\rho_{\rm de}
=
3\gamma M_{\rm p}^{2}
\left(
\dot{H}+2H^{2}
\right),
\end{equation}
with $\gamma$ a dimensionless parameter that quantifies the strength of the curvature contribution to dark energy.

Different from HDE and ADE, the dimensionless Hubble parameter of RDE has an analytical solution,
\begin{align}
E_{\rm RDE}^{2}(z)
=
&\frac{2\Omega_{m0}}{2-\gamma}(1+z)^3
+
\Omega_{r0}(1+z)^4
\nonumber\\
&+
\left(
1-\Omega_{r0}
-\frac{2\Omega_{m0}}{2-\gamma}
\right)
(1+z)^{4-2/\gamma},
\label{eq:E_RDE}
\end{align}
which automatically satisfies the normalization condition $E(0)=1$. The dark energy density parameter can be reconstructed through the cosmic energy partition:
\begin{equation}
\Omega_{\rm de}(z)
=
1
-
\frac{\Omega_{m0}(1+z)^3+\Omega_{r0}(1+z)^4}{E_{\rm RDE}^{2}(z)}.
\end{equation}
The parameter $\gamma$ controls the scaling behavior of RDE and shapes the late-time cosmic expansion history.

Using the conservation equation
\begin{equation}
\dot{\rho}_{\rm de}=-3H(1+w_{\rm de})\rho_{\rm de},
\end{equation}
the RDE equation-of-state parameter can be written as
\begin{equation}
w_{\rm de}(z)
=
-1+
\frac{1+z}{3X_{\rm de}(z)}
\frac{{\rm d}X_{\rm de}(z)}{{\rm d}z},
\label{eq:w_RDE_general}
\end{equation}
where $X_{\rm de}(z)\equiv \rho_{\rm de}(z)/\rho_{c0}$. From Eq.~\eqref{eq:E_RDE}, one obtains
\begin{equation}
X_{\rm de}(z)=A(1+z)^3+B(1+z)^p,
\end{equation}
with
\begin{equation}
A=\frac{\gamma\Omega_{m0}}{2-\gamma},\quad
B=1-\Omega_{r0}-\frac{2\Omega_{m0}}{2-\gamma},\quad
p=4-\frac{2}{\gamma}.
\end{equation}
Therefore,
\begin{equation}
w_{\rm de}(z)
=
-1+
\frac{
3A(1+z)^3+pB(1+z)^p
}{
3\left[A(1+z)^3+B(1+z)^p\right]
}.
\label{eq:w_RDE}
\end{equation}
This form shows explicitly that $\gamma$ controls both the expansion history and the dynamical behavior of RDE.

\section{Observational Data and Statistical Methodology}
\subsection{Cosmic Chronometer $H(z)$ Data}

Cosmic chronometer (CC) measurements provide a direct probe of the Hubble expansion rate by using the differential age evolution of massive, passively evolving galaxies. This method was first proposed by \citet{2002ApJ...573...37J} and relies on the relation
\begin{equation}
H(z)=-\frac{1}{1+z}\frac{dz}{dt},
\end{equation}
where $dz/dt$ is inferred from the relative age difference of old galaxy populations separated by a small redshift interval. Since the CC method does not rely on a cosmological distance ladder or standard candles, it provides a nearly model-independent measurement of the late-time expansion history.

In practice, the accuracy of CC measurements is limited not only by statistical uncertainties but also by several important systematic effects. As discussed in detail by \citet{2020ApJ...898...82M,2022LRR....25....6M}, the dominant sources of systematics arise from stellar population synthesis (SPS) modeling and the assumed initial mass function (IMF). These effects can introduce additional redshift-dependent uncertainties and correlations among different $H(z)$ measurements. Therefore, a reliable cosmological analysis should incorporate the full covariance structure rather than treating all data points as statistically independent.

In this work, we adopt the compilation of 31 CC measurements. These measurements are based on the observational data originally presented in \citep{2003ApJ...593..622J,2012JCAP...07..053M,2015MNRAS.450L..16M,2016JCAP...05..014M} and cover the redshift range $0.07\leq z\leq1.965$. Following the methodology of \citet{2020ApJ...898...82M}, we include the full systematic covariance matrix associated with SPS and IMF uncertainties. This treatment increases the effective uncertainty of the CC data compared with a purely statistical analysis and provides a more conservative and robust constraint on the background expansion history.

For a given cosmological model, the theoretical prediction $H_{\rm th}(z_i)$ is compared with the observed value $H_{\rm obs}(z_i)$ through
\begin{equation}
\chi_{H(z)}^2
=
\Delta \mathbf{H}^{\rm T}
\mathbf{C}_{H}^{-1}
\Delta \mathbf{H},
\end{equation}
where
\begin{equation}
\Delta H_i
=
H_{\rm obs}(z_i)-H_{\rm th}(z_i),
\end{equation}
and $\mathbf{C}_{H}$ is the full CC covariance matrix including both statistical and systematic contributions. The CC data directly constrain the absolute expansion rate and play a complementary role to SNe Ia and BAO measurements, helping to break degeneracies among $H_0$, $\Omega_{m0}$, and model-specific dark energy parameters.

\subsection{Type Ia Supernova Data}
Type Ia supernovae (SNe Ia) serve as standardizable candles and provide precise measurements of the luminosity-distance--redshift relation. In this work, we employ three recent SNe Ia compilations to constrain the late-time expansion history: Pantheon+, DES-Dovekie, and DESY5. These data sets cover complementary redshift ranges and calibration strategies, and are combined with cosmic chronometer $H(z)$ measurements and DESI DR2 BAO data to test the background evolution predicted by different holographic-inspired dark energy models.

We first use the Pantheon+ compilation \citep{2022ApJ...938..113S}, which contains 1701 high-quality light curves from 1590 spectroscopically confirmed SNe Ia over the redshift range $0.01<z<2.26$. To reduce the impact of low-redshift calibration systematics, we exclude the calibration subsample at $z<0.01$ and use the corresponding Hubble-diagram data together with the full covariance matrix\footnote{\url{https://github.com/PantheonPlusSH0ES/DataRelease}}, which includes both statistical uncertainties and systematic correlations. The resulting Pantheon+ sample provides a robust distance-redshift relation for cosmological inference.

We also include the fully recalibrated DES-Dovekie SNe Ia sample \citep{DES:2025sig}, which is based on a comprehensive reanalysis of the Dark Energy Survey (DES) 5-year supernova data. This data set incorporates several calibration and modeling improvements, including enhanced photometric cross-calibration, updated white-dwarf calibration observations across surveys, a retrained SALT3 light-curve model, and a corrected host-galaxy color law \citep{2024ApJ...973L..14D}. The final DES-Dovekie sample consists of approximately 1600 likely SNe Ia from DES together with about 200 low-redshift SNe Ia from external surveys, providing an updated and homogeneous basis for late-time cosmological constraints.

Finally, we consider the DES 5-year SNe Ia compilation, denoted as DESY5 \citep{2024ApJ...973L..14D}. This sample contains 1635 photometrically classified DES SNe Ia spanning $0.1<z<1.13$, supplemented by 194 low-redshift SNe Ia. The combined catalog includes 1829 objects\footnote{\url{https://github.com/des-science/DES-SN5YR}} and provides a uniformly selected data set with improved calibration and systematic-error control. Throughout this work, the three SNe Ia data sets are labeled as Pantheon+, DES-Dovekie, and DESY5, respectively.

For each supernova, the theoretical distance modulus is
\begin{equation}
\mu_{\rm th}(z)=5\log_{10}\left[\frac{D_L(z)}{\rm Mpc}\right]+25,
\end{equation}
where the luminosity distance in a flat FRW universe is
\begin{equation}
D_L(z)=(1+z)\clight\int_0^z \frac{dz'}{H(z')}.
\end{equation}
Since uncalibrated SNe Ia constrain relative distances, the absolute magnitude or an equivalent additive offset is treated as a nuisance parameter and marginalized over in the likelihood analysis. The supernova likelihood is constructed using the statistical and systematic covariance matrix:
\begin{equation}
\chi_{\rm SN}^2 =
\Delta \boldsymbol{\mu}^{\rm T}
\mathbf{C}_{\rm SN}^{-1}
\Delta \boldsymbol{\mu},
\end{equation}
with
\begin{equation}
\Delta \boldsymbol{\mu}
=
\boldsymbol{\mu}_{\rm obs}
-
\boldsymbol{\mu}_{\rm th}.
\end{equation}
The SNe Ia data provide high-precision constraints on the shape of the late-time distance-redshift relation and are particularly useful for testing deviations from $\LCDM$ expansion histories. When combined with $H(z)$ and BAO data, they help constrain dark energy dynamics while reducing degeneracies among $H_0$, $\Omega_{m0}$, and model-specific parameters.

\subsection{Dark Energy Spectroscopic Instrument DR2 BAO Data}

BAO provide a standard ruler set by the sound horizon at the baryon drag epoch, $r_d$, and represent one of the most robust geometrical probes of cosmic expansion. We adopt DESI DR2 BAO data \cite{Abdul2025}, which include galaxy, quasar, and Lyman-$\alpha$ forest tracers over a wide redshift range. Compared with previous measurements, DESI R2 provides improved distance precision and stronger constraining power for late-time dark energy models. 

The BAO observables constrain combinations of the transverse comoving distance, the Hubble distance, and the volume-averaged distance:
\begin{equation}
D_M(z)=\clight\int_0^z\frac{dz'}{H(z')},
\end{equation}
\begin{equation}
D_H(z)=\frac{\clight}{H(z)},
\end{equation}
and
\begin{equation}
D_V(z)=\left[zD_M^2(z)D_H(z)\right]^{1/3}.
\end{equation}
Depending on the tracer and redshift bin, measurements are provided either in anisotropic form through $D_M(z)/r_d$ and $D_H(z)/r_d$, or in isotropic form through $D_V(z)/r_d$. In our analysis, the sound horizon $r_d$ is treated as a free parameter rather than being fixed to a CMB-inferred value, allowing a late-time analysis without imposing an early-universe sound-horizon prior.

For each model, the BAO contribution to the total likelihood is
\begin{equation}
\chi_{\rm BAO}^2 =
\Delta \mathbf{B}^{\rm T}
\mathbf{C}_{\rm BAO}^{-1}
\Delta \mathbf{B},
\end{equation}
where $\mathbf{C}_{\rm BAO}$ is the covariance matrix and
\begin{equation}
\Delta \mathbf{B}=\mathbf{B}_{\rm obs}-\mathbf{B}_{\rm th}.
\end{equation}
Here $\mathbf{B}$ denotes the relevant distance vector, including $D_M/r_d$, $D_H/r_d$, or $D_V/r_d$ depending on the specific measurement. Due to its sensitivity to both transverse and line-of-sight distances, DESI DR2 is essential for breaking degeneracies between matter density, Hubble constant, and dynamical dark energy parameters.

\subsection{Redshift-Space Distortion Data}
 {To address the growth-level implications of the background-preferred solutions, we also include an extended set of redshift-space distortion measurements of $f\sigma_8(z)$. These data provide a direct probe of the product of the linear growth rate $f\equiv d\ln D/d\ln a$ and the amplitude of matter fluctuations $\sigma_8(z)$, and are therefore complementary to the background distance probes used above \cite{Kaiser:1987qv,Hamilton:1997zq,Song:2008qt}. We adopt an extended RSD compilation including low- and intermediate-redshift measurements from 6dFGS, SDSS MGS, WiggleZ, VIPERS, and FastSound, together with completed SDSS-IV/eBOSS DR16 measurements for LRG, QSO, and ELG samples \cite{Beutler:2012px,Howlett:2015wda,Blake:2011rj,Pezzotta:2016gbo,Okumura:2015lvp,GilMarin:2020bct,Hou:2020rse,Neveux:2020voa,Bautista:2020ahg,Alam:2020sor}. Compared with a compressed eBOSS-only subset, this extended compilation provides broader redshift leverage and a more direct phenomenological consistency check of the growth history, although it remains heterogeneous and is not equivalent to a full perturbation-level analysis.}

 {For a given background cosmology, the growth factor $D(a)$ is computed in the standard general-relativistic linear growth approximation,}
\begin{equation}
 {\frac{d^2D}{d\ln a^2}+\left[2+\frac{d\ln H}{d\ln a}\right]\frac{dD}{d\ln a}-\frac{3}{2}\Omega_m(a)D=0.}
\end{equation}
 {We normalize $D(a=1)=1$ and compare the theoretical prediction $f\sigma_8(z)=\sigma_{8,0}f(z)D(z)$ with the observed RSD vector. The present-day fluctuation amplitude $\sigma_{8,0}$ is treated as a free nuisance parameter only for the RSD-included data combination. The RSD contribution to the likelihood is}
\begin{equation}
 {\chi_{\rm RSD}^2=\Delta\mathbf{F}^{\rm T}\mathbf{C}_{\rm RSD}^{-1}\Delta\mathbf{F},}
\end{equation}
 {where $\Delta F_i=f\sigma_{8,{\rm obs}}(z_i)-f\sigma_{8,{\rm th}}(z_i)$. This treatment provides a phenomenological growth consistency test. It should not be interpreted as a full perturbation-level stability analysis of the dark energy models, especially for RDE, whose perturbative behavior depends on the full dynamical trajectory and perturbation sector.}

\subsection{Markov Chain Monte Carlo Analysis and Model Comparison}
We perform Bayesian cosmological parameter estimation using MCMC sampling. For each model and data combination, we generate posterior samples and check convergence through chain stability and autocorrelation behavior. The flat prior ranges for cosmological and model parameters are listed in Table \ref{tab:priors}.

\begin{table}[htbp]
\centering
\caption{Flat priors adopted in the MCMC parameter analysis.}
\label{tab:priors}
\renewcommand{\arraystretch}{1.4}
\begin{tabular}{cc}
\hline
Parameter & Prior Range \\
\hline
$H_0$          & $[50,\ 90]$ \\
$\Omega_{m0}$  & $[0.1,\ 0.5]$ \\
$c$            & $[0.1,\ 3]$ \\
$n$            & $[0.1,\ 5]$ \\
$\gamma$       & $[0.01,\ 1]$ \\
$r_d$          & $[130,\ 160]$  \\
 {$\sigma_{8,0}$} &  {$[0.4,\ 1.2]$} \\
\hline
\end{tabular}
\end{table}

All prior ranges are chosen to cover the relevant posterior region. The total joint likelihood for combined data sets is
\be
 {\chi^2_{\text{tot}} = \chi^2_{H(z)} + \chi^2_{\text{SN}} + \chi^2_{\text{BAO}} + \chi^2_{\text{RSD}}.}
\ee

To quantitatively evaluate the relative performance of different dark energy models, we use Akaike information criterion \cite{1985ITASS..33..387W} and Bayes information criterion, defined as
\be
\text{AIC} = \chi^2_{\text{min}} + 2k, \qquad
\text{BIC} = \chi^2_{\text{min}} + k \ln N,
\ee
where $k$ is the number of free model parameters and $N$ is the total number of observational data points. The relative statistical preference is quantified by
\[
\Delta \mathrm{AIC} = \mathrm{AIC}_{\mathrm{model}} - \mathrm{AIC}_{\Lambda\mathrm{CDM}},
\]
with the same definition for $\Delta \mathrm{BIC}$. We follow the standard interpretation that $|\Delta \mathrm{AIC}| < 2$ indicates statistically comparable models, while larger values indicate increasing preference for one model over the other.

Notably, information criteria reflect relative statistical fitting quality and parameter penalty; they do not by themselves establish the physical consistency of a cosmological model.

\section{Results}
In this section, we present posterior constraints on HDE, ADE, and RDE from five data combinations adopted in this work: $H(z)$+Pantheon+, $H(z)$+DESI DR2+Pantheon+, $H(z)$+DESI DR2+DES-Dovekie, $H(z)$+DESI DR2+DESY5, and  {$H(z)$+DESI DR2+DES-Dovekie+RSD}. The main cosmological parameters, model parameters, $1\sigma$ uncertainties,  {$r_d$, $\sigma_{8,0}$ where applicable}, and model comparison indices $\Delta$AIC and $\Delta$BIC are listed in Table~\ref{tab:summary}. We interpret the physical parameter evolution of each model, compare constrained parameters with the SH0ES local measurement ($H_0 = 73.17\pm0.86~\Mpc$) and Planck 2018 CMB results ($H_0 = 67.4\pm0.5~\Mpc$, $\Omega_{m0}=0.315\pm0.007$), and discuss the role of each observational probe in breaking parameter degeneracies.

\subsection{Constraints on the Holographic Dark Energy Model}

\begin{figure}[htbp]
    \centering
    \includegraphics[width=\columnwidth]{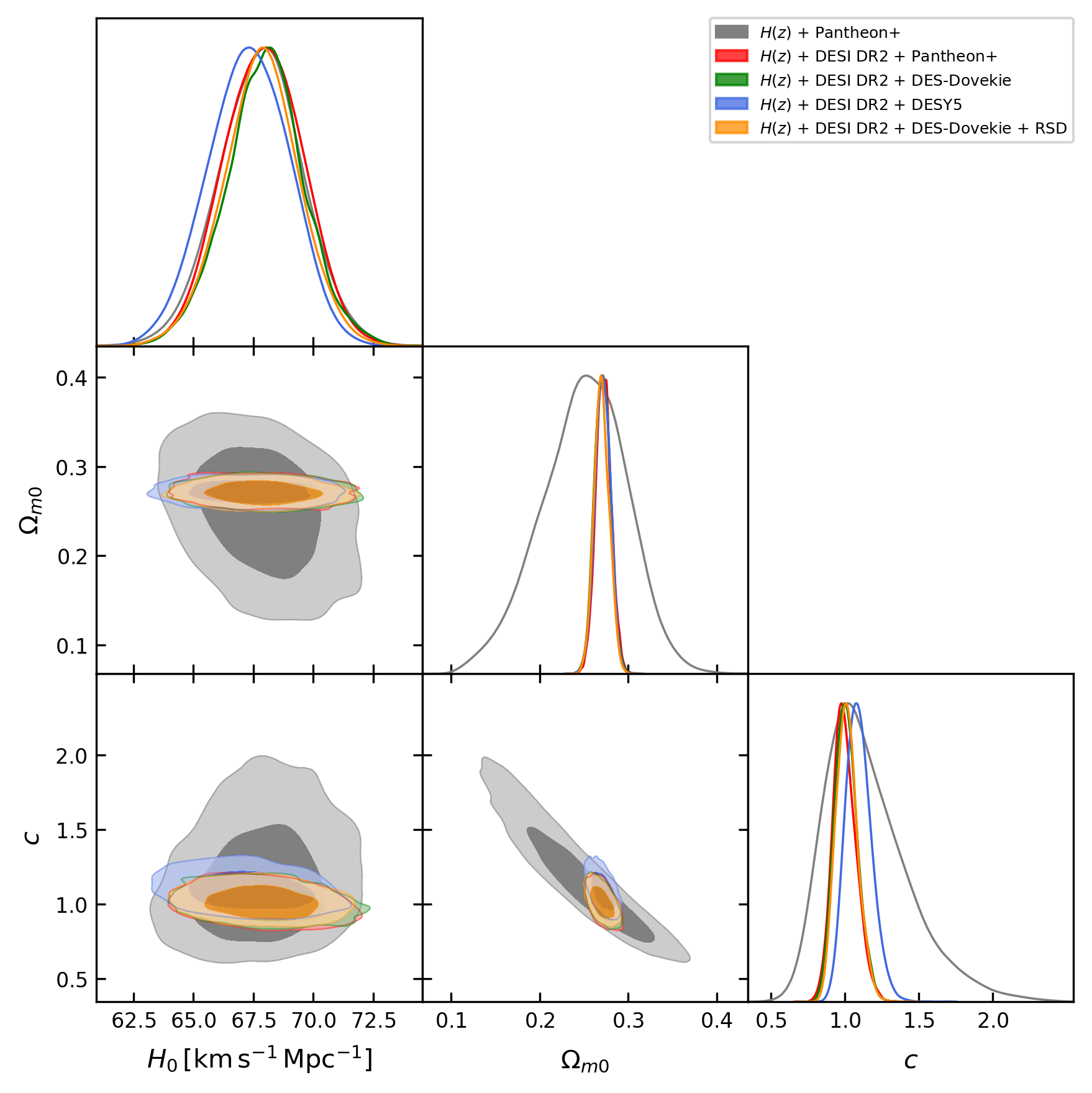}
    \caption{Confidence contours at $1\sigma$ and $2\sigma$ confidence levels for the HDE model from the  {five data combinations}.}
    \label{fig:hde}
\end{figure}

Figure~\ref{fig:hde} shows the $1\sigma$ and $2\sigma$ confidence contours of HDE for the five data combinations. The corresponding posterior constraints are summarized in Table~\ref{tab:summary}. For the first four data combinations, the constraints remain consistent with the background-only results discussed above. For the $H(z)$+DESI DR2+DES-Dovekie+RSD combination, we obtain $H_0=67.860^{+1.534}_{-1.579}~\Mpc$, $\Omega_{m0}=0.270^{+0.009}_{-0.008}$, $c=1.011^{+0.076}_{-0.072}$,  {$r_d=146.957^{+3.419}_{-3.204}$ Mpc, and $\sigma_{8,0}=0.875^{+0.037}_{-0.039}$.}

The holographic parameter $c$ changes from slightly above unity in the $H(z)$+Pantheon+ combination to very close to unity after adding BAO and RSD information.  {Since $c$ fixes the normalization of the holographic relation between the dark-energy density and the future event-horizon infrared cutoff, its convergence toward unity has a direct physical meaning rather than being only a numerical fitting trend.} This indicates that the preferred HDE evolution approaches the de Sitter boundary in the future dark-energy-dominated limit. Although values slightly below or above unity are allowed within $1\sigma$, the current data do not support a strong phantom regime.  {In other words, the data select the boundary region between phantom-like and quintessence-like HDE behavior, where the background expansion becomes close to that of a cosmological constant while still retaining the dynamical relation between $w_{\rm de}(z)$, $\Omega_{\rm de}(z)$, and $c$.} The constrained $H_0$ values of HDE cluster steadily around $67$--$68~\Mpc$, close to the Planck CMB inference and far below the SH0ES local measurement. Therefore, HDE does not significantly alleviate the Hubble tension in this analysis. The matter density constrained by the full DESI DR2+SNe Ia combinations converges to $\Omega_{m0}\simeq0.270$--$0.272$.  {This moderate lowering of $\Omega_{m0}$ relative to Planck reflects the usual distance-probe degeneracy between matter density and dark-energy dynamics, but it is not large enough to move the model into a qualitatively different expansion regime.} The AIC differences relative to $\LCDM$ are $\Delta\text{AIC}=1.07,-1.66,-2.48,-7.03$, and  {$-3.76$ for the RSD-included data set}. However, the corresponding BIC values show that the statistical evidence remains model-selection dependent, with  {the RSD-included case giving $\Delta\text{BIC}=1.78$}.  {Thus, the AIC improvement should be read as an improved best-fit description allowed by the extra holographic degree of freedom, whereas the BIC result suggests that, after penalizing the additional model complexity, current data do not yet provide a compelling need for the horizon-scale mechanism.}

Our HDE constraints are consistent with recent late-time DESI-DR2 analyses. For example, a recent study of HDE and RDE using CC+GRB+DESI-DR2+DES-Dovekie obtained $H_0=68.2\pm2.8~\Mpc$, $\Omega_{m0}=0.272\pm0.009$, and $c=1.000^{+0.064}_{-0.082}$ \cite{2026MNRAS.547ag365N}.  {At the background level, this near-de Sitter HDE solution is increasingly degenerate with $\LCDM$. Perturbation-level observables, such as RSD, weak lensing, galaxy clustering, and the ISW effect, are therefore needed to test whether the dynamical relation between $w_{\rm de}(z)$, $\Omega_{\rm de}(z)$, and $c$ can be distinguished from a cosmological constant.}

\subsection{Constraints on the Agegraphic Dark Energy Model}

\begin{figure}[htbp]
    \centering
    \includegraphics[width=\columnwidth]{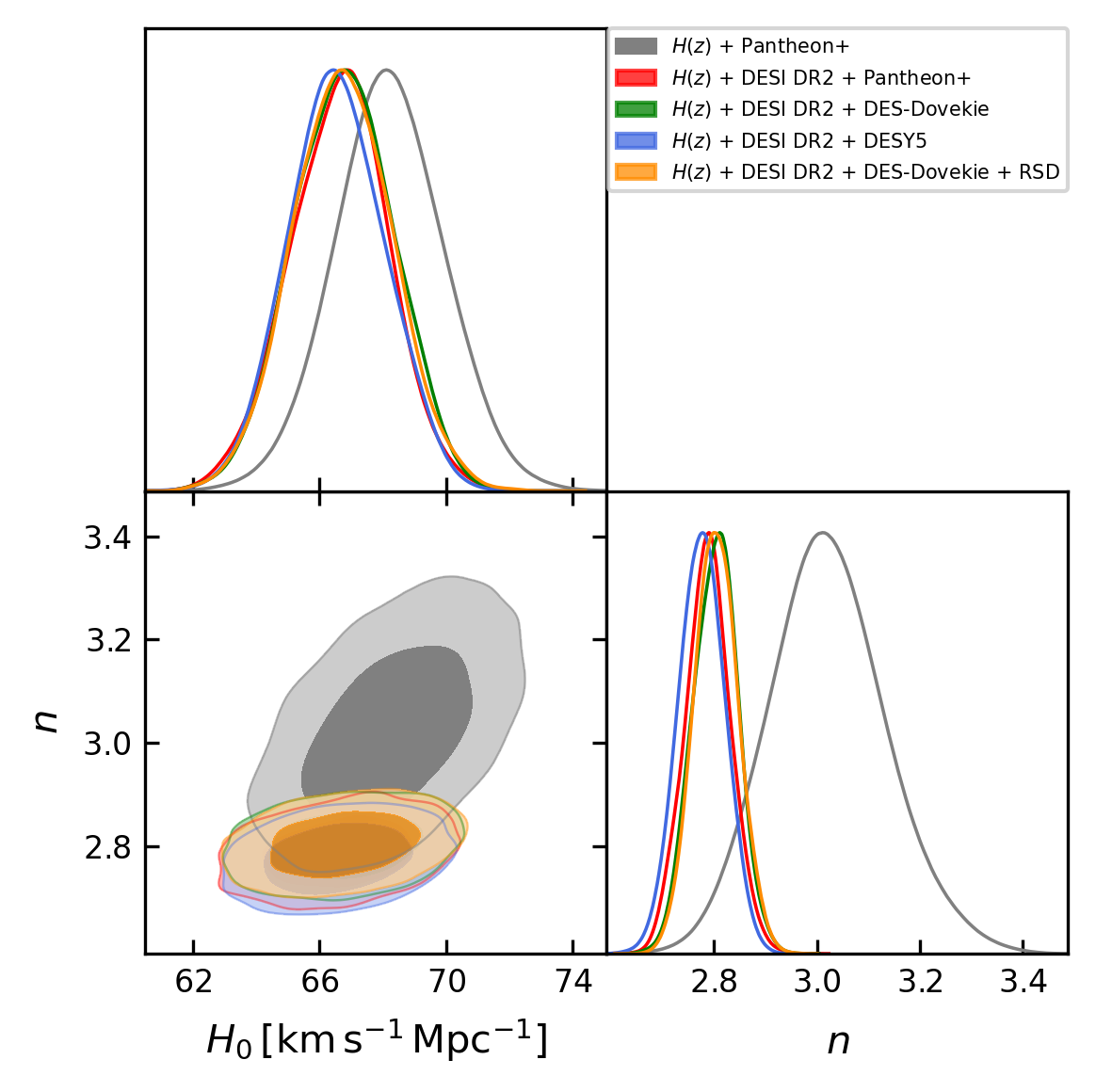}
    \caption{Same as Fig.~\ref{fig:hde}, but for the ADE model.}
    \label{fig:ade}
\end{figure}

Figure~\ref{fig:ade} shows the confidence contours of ADE from different data combinations, with posterior constraints listed in Table~\ref{tab:summary}. For the first four data combinations, the constraints remain consistent with the background-only results discussed above. For the $H(z)$+DESI DR2+DES-Dovekie+RSD combination, we obtain $H_0=66.734^{+1.591}_{-1.579}~\Mpc$, $n=2.804^{+0.042}_{-0.042}$, { $r_d=147.075^{+3.529}_{-3.323}$ Mpc, and $\sigma_{8,0}=0.893^{+0.035}_{-0.036}$.}

Different from HDE and RDE, ADE does not treat $\Omega_{m0}$ as an independent free parameter; it is fixed by the flatness and normalization conditions for each pair of $H_0$ and $n$.  {This makes ADE more predictive than models in which the matter density can be shifted independently, because the matter fraction, the expansion history, and the dark-energy equation of state are tied together by the conformal-age prescription.} The constrained $H_0$ values prefer $66.5$--$68.2~\Mpc$, close to or slightly below the Planck value and clearly below the SH0ES local calibration.  {The low $H_0$ preference therefore has a physical origin: the agegraphic scaling does not provide enough late-time freedom to raise the recent expansion rate while simultaneously preserving the BAO and SNe Ia distance relations.} The model parameter $n$ converges stably around $2.8$ across the BAO- and RSD-included data combinations, showing internal parameter stability.  {Since $n$ controls the strength of the dark-energy density associated with the conformal age of the Universe, this stability indicates that the data consistently select nearly the same age scale for the onset of late-time acceleration.} The information criteria, however, generally disfavor ADE relative to $\LCDM$ once DESI DR2 BAO data are included,  {with the RSD-included combination giving $\Delta\text{AIC}=\Delta\text{BIC}=3.51$. This suggests that the physical economy of ADE also limits its fitting flexibility: the conformal-age mechanism remains observationally allowed, but current BAO-included data do not require it over $\LCDM$.}

The value of $n$ inferred here is consistent with the standard ADE benchmark range.  {The corresponding reconstructed $w_{\rm de}(z)$ remains quintessence-like, so ADE represents a restricted dynamical dark-energy scenario rather than a flexible empirical $w(a)$ fit.} We note that direct observational constraints on the minimal ADE model are mostly from earlier studies, while recent ADE works tend to focus on entropy-corrected, interacting, or modified-gravity extensions whose parameters are not directly comparable with the standard ADE parameter $n$.

\subsection{Constraints on the Ricci Dark Energy Model}

\begin{figure}[htbp]
    \centering
    \includegraphics[width=\columnwidth]{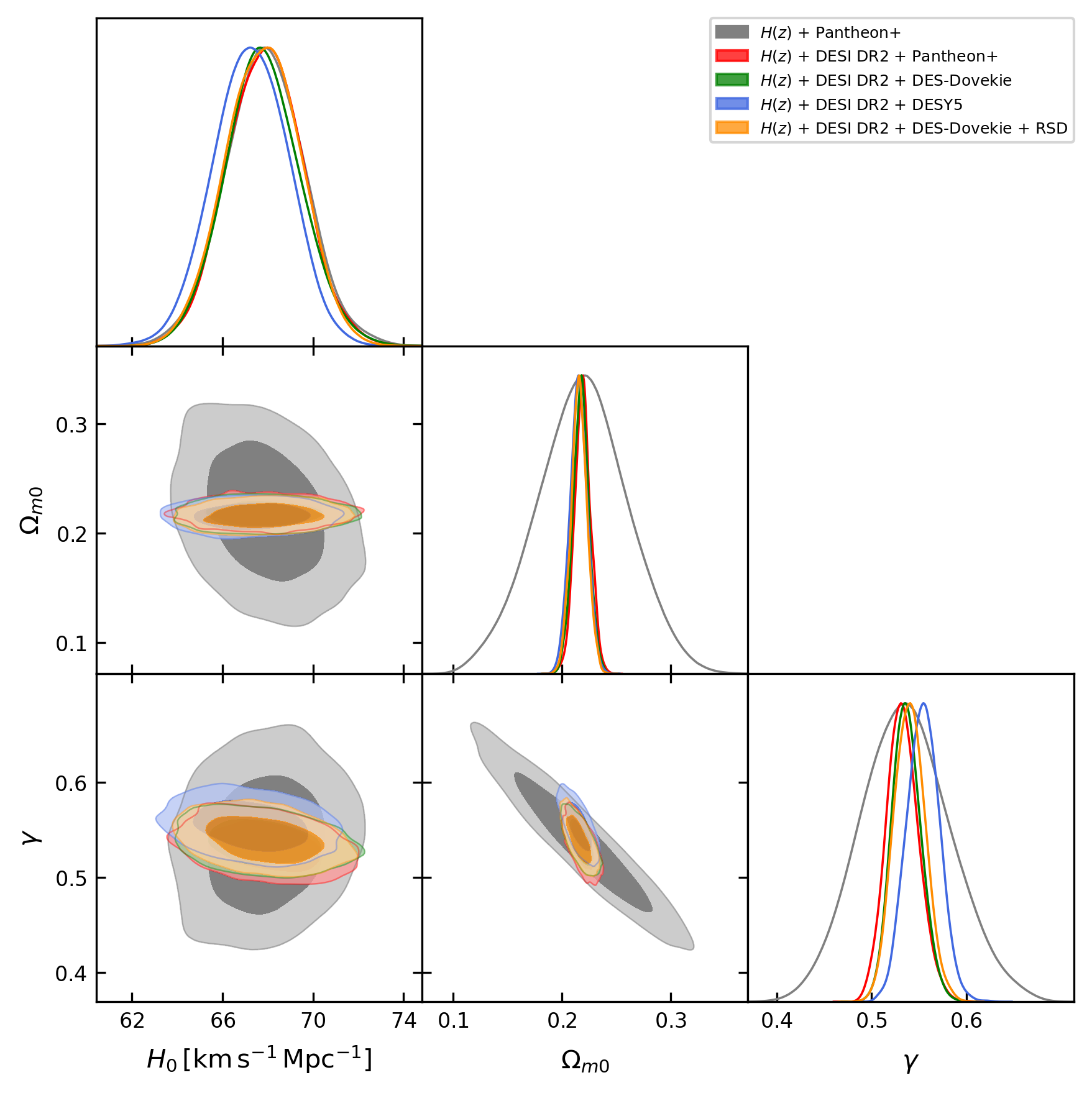}
    \caption{Same as Fig.~\ref{fig:hde}, but for the RDE model.}
    \label{fig:rde}
\end{figure}

\begin{figure*}[t]
    \centering
    \includegraphics[width=\textwidth]{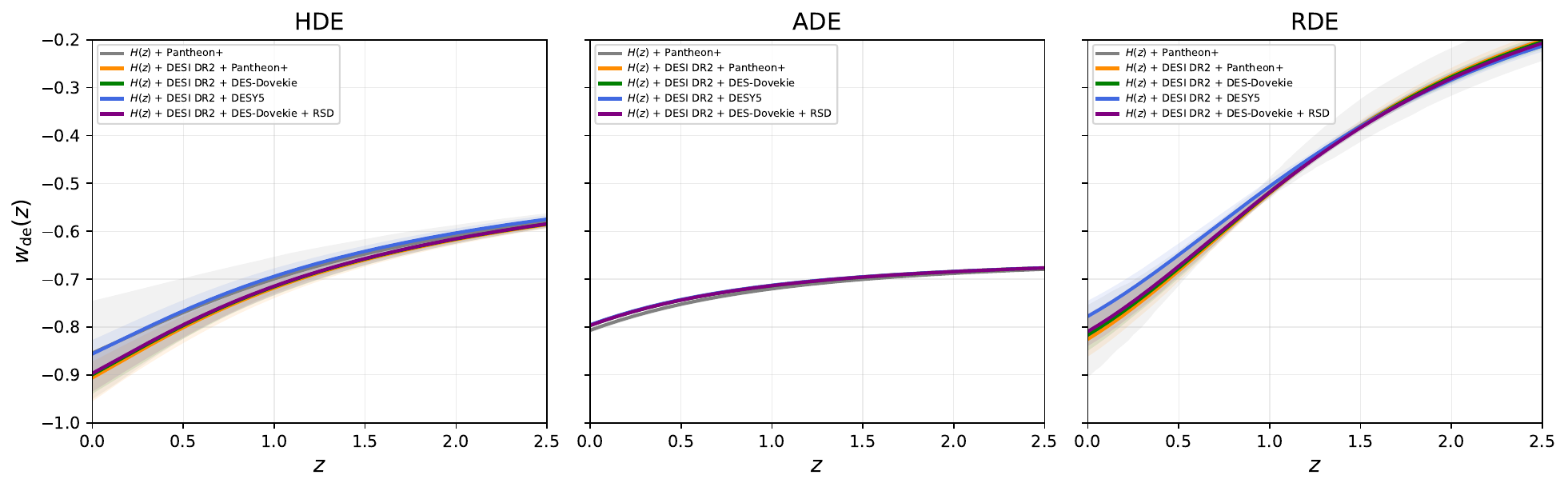}
    \caption{Redshift evolution of the dark energy equation-of-state parameter $w_{\rm de}(z)$ for the three holographic-inspired dark energy models. From left to right, the panels correspond to HDE, ADE, and RDE, respectively. Solid curves show posterior medians reconstructed from the MCMC samples, and shaded regions denote the $1\sigma$ credible intervals.  {The same axis ranges are used in all three panels.}}
    \label{fig:wz_evolution}
\end{figure*}

The confidence contours of RDE are presented in Fig.~\ref{fig:rde}, with posterior constraints summarized in Table~\ref{tab:summary}. For the first four data combinations, the constraints remain consistent with the background-only results discussed above. For the $H(z)$+DESI DR2+DES-Dovekie+RSD combination, we obtain $H_0=67.822^{+1.655}_{-1.691}~\Mpc$, $\Omega_{m0}=0.216^{+0.007}_{-0.007}$, $\gamma=0.540^{+0.017}_{-0.017}$,  {$r_d=146.988^{+3.563}_{-3.344}$ Mpc, and $\sigma_{8,0}=0.961^{+0.043}_{-0.041}$.}

Across all data combinations, the matter density of RDE remains persistently low, $\Omega_{m0}\approx0.215$--$0.219$, substantially below the Planck 2018 value $\Omega_{m0}=0.315\pm0.007$. Quantifying this comparison using the BAO-included uncertainties gives a tension of roughly $9\sigma$ with the Planck $\LCDM$-inferred value, although this should not be interpreted as a full CMB constraint within the RDE model.  {The origin of this preference is tied to the RDE construction itself: because the dark-energy density is linked to the Ricci scalar, part of the late-time distance behavior normally attributed to the matter component and a cosmological constant can be absorbed by the curvature-related dark-energy term. The model can therefore fit background distances with a lower matter fraction, but this comes at the cost of moving away from the matter density favored by early-universe and standard structure-formation analyses.} This low matter density may indicate a potential tension with early-universe and structure-formation constraints, but such a conclusion cannot be established from background data alone.  {The RSD-included fit provides a first growth-level consistency check and yields $\sigma_{8,0}\simeq0.96$. The combination of low $\Omega_{m0}$ and relatively high $\sigma_{8,0}$ is physically informative: within the simplified growth treatment adopted here, a larger present-day fluctuation amplitude is needed to match the observed $f\sigma_8(z)$ data when the background matter density is low. However, this compensation is not a proof of perturbative viability, and a full perturbation analysis of RDE is still required. Previous perturbative studies have shown that stability depends on the full background trajectory and on derivatives of $H(z)$, not only on present-day effective quantities \cite{2013PhRvD..87l3002D}.} {Therefore, although the preferred RDE solution is non-phantom-like over the plotted late-time range, this is not sufficient to certify that it belongs to the perturbatively stable region identified in Ref.~\cite{2013PhRvD..87l3002D}. A direct stability assessment would require solving the linear perturbation system for the full best-fit background trajectory, which is beyond the scope of the present late-time parameter-constraint analysis.}

The model comparison gives $\Delta\text{AIC}=1.06,-2.81,-4.18,-9.81$, and  {$-6.38$ for the RSD-included combination}. Negative $\Delta\text{AIC}$ values appear in the BAO-included cases, suggesting an improvement in best-fit quality relative to $\LCDM$. Nevertheless, the corresponding $\Delta\text{BIC}$ values are less favorable and  {become negative for DESY5 and mildly negative for the RSD-included combination. Thus, the AIC improvement should be interpreted cautiously: RDE can fit the late-time distances well because its curvature-linked density has more background-evolution structure than a constant vacuum energy, but the same mechanism drives the model toward a low-$\Omega_{m0}$ region. Its statistical competitiveness therefore does not by itself establish physical viability; the low matter density, high fitted $\sigma_{8,0}$, and perturbative consistency remain important caveats.}

\begin{table*}[htbp]
\centering
\renewcommand{\arraystretch}{1.55}
\setlength{\tabcolsep}{3.0pt}
\caption{Posterior median values with $1\sigma$ uncertainties for HDE, ADE, RDE, and  {$\LCDM$} under the five data combinations.  {The corresponding $\LCDM$ posterior constraints are included for the same five data combinations. The sound horizon $r_d$ is reported for BAO-included combinations, and $\sigma_{8,0}$ is reported for the RSD-included combination.} $\Delta$AIC and $\Delta$BIC are calculated relative to the corresponding $\LCDM$ baseline for the same data combination. The unit of $H_0$ is km s$^{-1}$ Mpc$^{-1}$, and  {the unit of $r_d$ is Mpc.}}
\label{tab:summary}
\resizebox{\textwidth}{!}{%
\begin{tabular}{ccccccccccc}
\hline
\hline
\textbf{Model} & \textbf{Dataset} & \textbf{$H_0$} & \textbf{$\Omega_m$} & \textbf{$c$} & \textbf{$n$} & \textbf{$\gamma$} &  {\textbf{$r_d$}} &  {\textbf{$\sigma_{8,0}$}} & \textbf{$\Delta$AIC} & \textbf{$\Delta$BIC}\\
\hline
 {$\LCDM$} &  {$H(z)$+Pantheon+} &  {$67.543^{+1.720}_{-1.644}$} &  {$0.332^{+0.018}_{-0.017}$} & -- & -- & -- & -- & -- &  {$0.00$} &  {$0.00$} \\
 {$\LCDM$} &  {$H(z)$+DR2+Pantheon+} &  {$68.783^{+1.579}_{-1.591}$} &  {$0.304^{+0.008}_{-0.008}$} & -- & -- & -- &  {$146.893^{+3.402}_{-3.347}$} & -- &  {$0.00$} &  {$0.00$} \\
 {$\LCDM$} &  {$H(z)$+DR2+DES-Dovekie} &  {$68.771^{+1.695}_{-1.684}$} &  {$0.306^{+0.007}_{-0.007}$} & -- & -- & -- &  {$146.596^{+3.657}_{-3.372}$} & -- &  {$0.00$} &  {$0.00$} \\
 {$\LCDM$} &  {$H(z)$+DR2+DESY5} &  {$68.524^{+1.658}_{-1.568}$} &  {$0.310^{+0.008}_{-0.008}$} & -- & -- & -- &  {$146.807^{+3.350}_{-3.456}$} & -- &  {$0.00$} &  {$0.00$} \\
 {$\LCDM$} &  {$H(z)$+DR2+DES-Dovekie+RSD} &  {$68.665^{+1.671}_{-1.586}$} &  {$0.305^{+0.008}_{-0.007}$} & -- & -- & -- &  {$146.989^{+3.316}_{-3.464}$} &  {$0.820^{+0.030}_{-0.033}$} &  {$0.00$} &  {$0.00$} \\
\hline
Holographic & $H(z)$+Pantheon+ & $67.850^{+1.671}_{-1.756}$ & $0.252^{+0.046}_{-0.051}$ & $1.107^{+0.326}_{-0.227}$ & -- & -- & -- & -- & $1.07$ & $6.46$ \\
Holographic & $H(z)$+DR2+Pantheon+ & $67.934^{+1.659}_{-1.695}$ & $0.272^{+0.009}_{-0.008}$ & $0.993^{+0.086}_{-0.069}$ & -- & -- &  {$146.858^{+3.421}_{-3.424}$} & -- & $-1.66$ & $3.74$ \\
Holographic & $H(z)$+DR2+DES-Dovekie & $68.022^{+1.540}_{-1.630}$ & $0.271^{+0.009}_{-0.009}$ & $1.005^{+0.081}_{-0.074}$ & -- & -- &  {$146.656^{+3.354}_{-3.339}$} & -- & $-2.48$ & $3.05$ \\
Holographic & $H(z)$+DR2+DESY5 & $67.326^{+1.701}_{-1.737}$ & $0.271^{+0.009}_{-0.009}$ & $1.089^{+0.093}_{-0.082}$ & -- & -- &  {$146.636^{+3.816}_{-3.406}$} & -- & $-7.03$ & $-1.50$ \\
Holographic &  {$H(z)$+DR2+DES-Dovekie+RSD} &  {$67.860^{+1.534}_{-1.579}$} &  {$0.270^{+0.009}_{-0.008}$} &  {$1.011^{+0.076}_{-0.072}$} & -- & -- &  {$146.957^{+3.419}_{-3.204}$} &  {$0.875^{+0.037}_{-0.039}$} &  {$-3.76$} &  {$1.78$} \\
\hline
Agegraphic & $H(z)$+Pantheon+ & $68.181^{+1.753}_{-1.698}$ & -- & -- & $3.018^{+0.116}_{-0.109}$ & -- & -- & -- & $0.09$ & $0.09$ \\
Agegraphic & $H(z)$+DR2+Pantheon+ & $66.686^{+1.462}_{-1.632}$ & -- & -- & $2.788^{+0.044}_{-0.044}$ & -- &  {$146.898^{+3.591}_{-3.132}$} & -- & $5.82$ & $5.82$ \\
Agegraphic & $H(z)$+DR2+DES-Dovekie & $66.764^{+1.614}_{-1.592}$ & -- & -- & $2.804^{+0.041}_{-0.046}$ & -- &  {$146.954^{+3.615}_{-3.460}$} & -- & $5.39$ & $5.39$ \\
Agegraphic & $H(z)$+DR2+DESY5 & $66.507^{+1.594}_{-1.511}$ & -- & -- & $2.776^{+0.045}_{-0.046}$ & -- &  {$147.018^{+3.381}_{-3.341}$} & -- & $-4.44$ & $-4.44$ \\
Agegraphic &  {$H(z)$+DR2+DES-Dovekie+RSD} &  {$66.734^{+1.591}_{-1.579}$} & -- & -- &  {$2.804^{+0.042}_{-0.042}$} & -- &  {$147.075^{+3.529}_{-3.323}$} &  {$0.893^{+0.035}_{-0.036}$} &  {$3.51$} &  {$3.51$} \\
\hline
Ricci & $H(z)$+Pantheon+ & $67.897^{+1.724}_{-1.730}$ & $0.219^{+0.041}_{-0.042}$ & -- & -- & $0.536^{+0.049}_{-0.047}$ & -- & -- & $1.06$ & $6.45$ \\
Ricci & $H(z)$+DR2+Pantheon+ & $67.899^{+1.645}_{-1.714}$ & $0.219^{+0.008}_{-0.007}$ & -- & -- & $0.532^{+0.018}_{-0.016}$ &  {$147.033^{+3.454}_{-3.326}$} & -- & $-2.81$ & $2.59$ \\
Ricci & $H(z)$+DR2+DES-Dovekie & $67.760^{+1.710}_{-1.588}$ & $0.218^{+0.007}_{-0.007}$ & -- & -- & $0.537^{+0.016}_{-0.015}$ &  {$147.130^{+3.383}_{-3.484}$} & -- & $-4.18$ & $1.35$ \\
Ricci & $H(z)$+DR2+DESY5 & $67.277^{+1.671}_{-1.669}$ & $0.215^{+0.008}_{-0.008}$ & -- & -- & $0.554^{+0.017}_{-0.018}$ &  {$146.986^{+3.384}_{-3.480}$} & -- & $-9.81$ & $-4.27$ \\
Ricci &  {$H(z)$+DR2+DES-Dovekie+RSD} &  {$67.822^{+1.655}_{-1.691}$} &  {$0.216^{+0.007}_{-0.007}$} & -- & -- &  {$0.540^{+0.017}_{-0.017}$} &  {$146.988^{+3.563}_{-3.344}$} &  {$0.961^{+0.043}_{-0.041}$} &  {$-6.38$} &  {$-0.85$} \\
\hline
\hline
\end{tabular}}
\end{table*}

Table~\ref{tab:summary} summarizes the main parameter constraints and information-criterion results for the three holographic-inspired models.  {It also includes the corresponding $\LCDM$ posterior constraints for the same five data combinations, so that both the parameter constraints and the $\Delta$AIC/$\Delta$BIC comparisons can be read directly from the table.} After DESI DR2 BAO data are included, HDE and RDE give similar values of $H_0\simeq67$--$68~\Mpc$, while ADE prefers a slightly lower value, $H_0\simeq66.5$--$66.8~\Mpc$. None of the three models significantly alleviates the Hubble tension. The HDE parameter is driven toward $c\simeq1$, indicating an evolution close to the de Sitter boundary, whereas ADE gives a stable $n\simeq2.8$ and RDE gives a stable $\gamma\simeq0.53$--$0.55$.  {These three parameters encode different physical mechanisms: $c$ measures the strength of the event-horizon infrared cutoff in HDE, $n$ fixes the conformal-age scale in ADE, and $\gamma$ controls the Ricci-curvature contribution in RDE. Their stability across BAO-included combinations therefore indicates that the data are not merely constraining generic dark-energy flexibility, but are selecting distinct physical regimes within each model.}  {The BAO-included constraints give $r_d\simeq146.6$--$147.2$ Mpc, and the RSD-included fit gives $\sigma_{8,0}=0.875^{+0.037}_{-0.039}$, $0.893^{+0.035}_{-0.036}$, and $0.961^{+0.043}_{-0.041}$ for HDE, ADE, and RDE, respectively.} The main physical weakness of RDE is its persistently low matter density, $\Omega_{m0}\simeq0.215$--$0.219$. From the information criteria, HDE is statistically competitive with $\LCDM$, ADE is generally disfavored once BAO data are included, and RDE improves AIC but must be interpreted with caution because of the low-$\Omega_{m0}$ preference and unresolved perturbative consistency.

\begin{table*}[htbp]
\centering
\renewcommand{\arraystretch}{1.35}
\setlength{\tabcolsep}{5.0pt}
\caption{ {Comparison between the constraints obtained in this work and those reported in Ref.~\cite{2026MNRAS.547ag365N} for the closest available HDE and RDE data combinations. The data combinations are not identical: our combination uses $H(z)+$DESI DR2+DES-Dovekie, whereas Ref.~\cite{2026MNRAS.547ag365N} uses CC+GRB+DESI-DR2+DES-Dovekie. The comparison is therefore intended as a consistency check for the overlapping DESI DR2+DES-Dovekie late-time constraints.}}
\label{tab:ref90_comparison}
\resizebox{\textwidth}{!}{%
\begin{tabular}{cccccc}
\hline
\hline
\textbf{Model} & \textbf{Data combination} & \textbf{Source} & \textbf{$H_0$} & \textbf{$\Omega_{m0}$} & \textbf{Model parameter} \\
\hline
 {HDE} &  {$H(z)+$DESI DR2+DES-Dovekie} &  {This work} &  {$68.022^{+1.540}_{-1.630}$} &  {$0.271^{+0.009}_{-0.009}$} &  {$c=1.005^{+0.081}_{-0.074}$} \\
 {HDE} &  {CC+GRB+DESI-DR2+DES-Dovekie} &  {Ref.~\cite{2026MNRAS.547ag365N}} &  {$68.2\pm2.8$} &  {$0.2721\pm0.0085$} &  {$c=0.999^{+0.064}_{-0.082}$} \\
 {RDE} &  {$H(z)+$DESI DR2+DES-Dovekie} &  {This work} &  {$67.760^{+1.710}_{-1.588}$} &  {$0.218^{+0.007}_{-0.007}$} &  {$\gamma=0.537^{+0.016}_{-0.015}$} \\
 {RDE} &  {CC+GRB+DESI-DR2+DES-Dovekie} &  {Ref.~\cite{2026MNRAS.547ag365N}} &  {$68.3\pm2.8$} &  {$0.2179\pm0.0078$} &  {$\gamma=0.536\pm0.016$} \\
\hline
\hline
\end{tabular}}
\end{table*}

 {Table~\ref{tab:ref90_comparison} shows that the differences between our results and Ref.~\cite{2026MNRAS.547ag365N} are not statistically significant for the closest overlapping data combinations. For HDE, the shifts in $H_0$, $\Omega_{m0}$, and $c$ are all much smaller than the combined $1\sigma$ uncertainties, confirming that both analyses select the same near-de Sitter region $c\simeq1$. For RDE, the agreement is even tighter: the preferred values of $\Omega_{m0}$ and $\gamma$ are nearly identical within the quoted uncertainties. Therefore, the main physical conclusion is robust: late-time DESI DR2+DES-Dovekie constraints drive HDE toward a near-$\LCDM$ background limit and RDE toward a low-$\Omega_{m0}$, $\gamma\simeq0.54$ solution.}

\begin{figure*}[t]
    \centering
    \includegraphics[width=\textwidth]{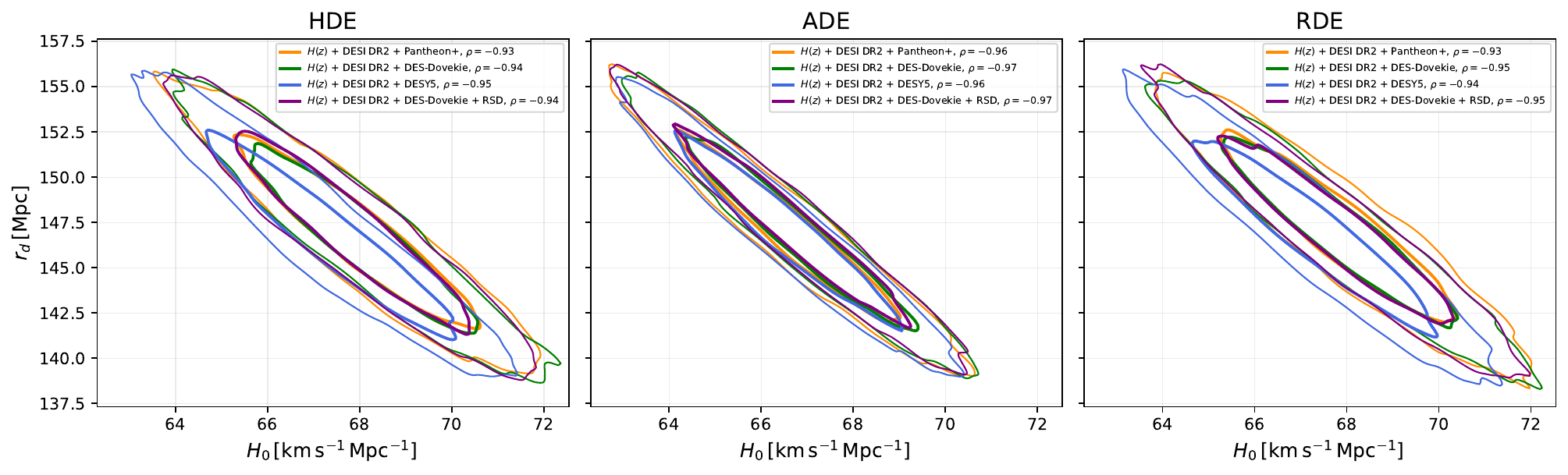}
    \caption{ {$H_0$--$r_d$ posterior contours for the BAO-included data combinations. From left to right, the panels correspond to HDE, ADE, and RDE, respectively. The strong negative correlations illustrate the degeneracy induced by treating the sound horizon $r_d$ as a free late-time parameter.}}
    \label{fig:h0_rd}
\end{figure*}

 {Because BAO measurements constrain ratios such as $D_M/r_d$, $D_H/r_d$, and $D_V/r_d$, treating $r_d$ as a free parameter naturally introduces a strong degeneracy with $H_0$ \cite{2015MNRAS.448.3463C,2020ApJ...904L..17P}. Figure~\ref{fig:h0_rd} shows the $H_0$--$r_d$ posterior contours for all BAO-included data combinations. The Pearson correlation coefficients are negative for all models and data combinations, typically $\rho(H_0,r_d)\simeq-0.93$ to $-0.97$. Therefore, the inferred $H_0$ values should be interpreted as late-time constraints marginalized over the sound-horizon scale, rather than CMB-calibrated determinations. The fact that $r_d$ remains close to the standard sound-horizon scale further indicates that these holographic-inspired models do not reduce the Hubble tension by shifting the BAO ruler to a substantially smaller value.}

\begin{figure}[htbp]
    \centering
    \includegraphics[width=\columnwidth]{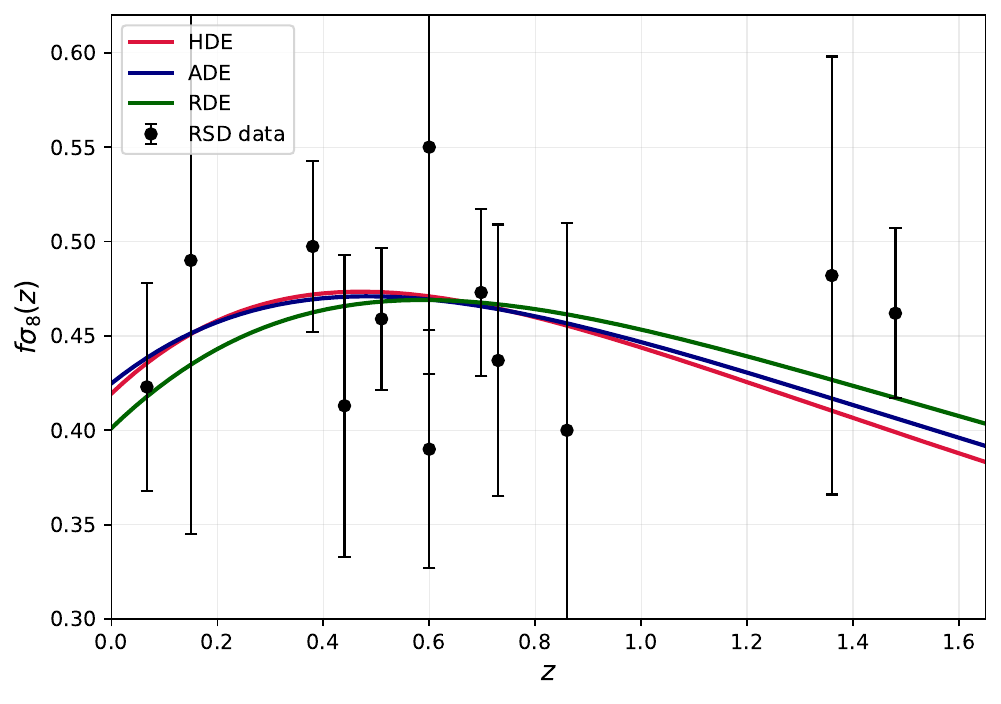}
    \caption{ {RSD measurements of $f\sigma_8(z)$ compared with the model predictions from the $H(z)$+DESI DR2+DES-Dovekie+RSD combination. The curves use the posterior median parameters for each model, including the fitted value of $\sigma_{8,0}$.}}
    \label{fig:rsd_fs8}
\end{figure}

 {Figure~\ref{fig:rsd_fs8} compares the RSD data with the predicted $f\sigma_8(z)$ curves for the three models using the RSD-included posterior medians. The current extended RSD compilation provides a broader redshift-leverage growth diagnostic than the compressed SDSS/eBOSS-only subset, although it remains heterogeneous and should not be interpreted as a complete perturbation-level stability test. The fitted $\sigma_{8,0}$ values should therefore be understood as the fluctuation amplitudes required for each background solution to match the adopted growth data, rather than as final predictions of a fully specified perturbation theory. In particular, the RDE curve remains compatible with the adopted RSD points after marginalizing over $\sigma_{8,0}$, but the relatively high $\sigma_{8,0}$ value and low $\Omega_{m0}$ preference still emphasize the need for a full perturbation-level treatment in future work.}

\subsection{Evolution of the Dark Energy Equation of State}

To further examine the dynamical behavior of the three models, we reconstruct the redshift evolution of the dark energy equation-of-state parameter $w_{\rm de}(z)$ using posterior samples from the five data combinations, as shown in Fig.~\ref{fig:wz_evolution}.  {The three panels use the same axis ranges, allowing the redshift evolution of the three models to be compared directly.} For HDE, $w_{\rm de}(z)$ remains above $-1$ over the plotted redshift range and has a present-day value around $w_0\simeq -0.9$, consistent with the fact that the BAO-included constraints drive $c$ close to unity while $\sqrt{\Omega_{\rm de0}}<1$.  {At the background level, this near-de Sitter region is increasingly similar to $\LCDM$, and perturbation-level observables are needed to break the degeneracy.} For ADE, the evolution is purely quintessence-like, as expected from Eq.~\eqref{eq:w_ADE}.  {This behavior follows from the conformal-age construction and is therefore more restricted than a freely chosen phenomenological $w(a)$.} For RDE, $w_{\rm de}(z)$ is reconstructed from Eq.~\eqref{eq:w_RDE}; it remains above $-1$ in the favored parameter region, with a present-day value around $w_0\simeq -0.8$ and a rapid increase toward higher redshift.  {The rapid redshift evolution reflects the explicit dependence of the RDE density on the curvature-related background terms, which makes RDE more flexible at the background level but also more sensitive to perturbative consistency.} Overall, the reconstructed equation-of-state evolution indicates that the present late-time data favor smooth quintessence-like or near-de Sitter behavior rather than a statistically significant transition across $w_{\rm de}=-1$.

 {To facilitate comparison with CPL-like descriptions, we also derive $w_0$ and an effective CPL slope $w_a^{\rm eff}$ from the MCMC posterior samples. For each posterior sample, $w_0$ is computed directly as $w_{\rm de}(z=0)$. We then fit the model-predicted $w_{\rm de}(a)$ over the same redshift range as Fig.~\ref{fig:wz_evolution}, $0\le z\le2.5$, to the CPL form $w(a)=w_0+w_a(1-a)$ while keeping $w_0$ fixed; the fitted slope is denoted $w_a^{\rm eff}$. These quantities are not independent CPL constraints and should not be interpreted as the result of a CPL model fit. They are low-redshift CPL projections of the corresponding holographic-inspired models, introduced only to make the direction and strength of the predicted equation-of-state evolution easier to compare with CPL-like DESI DR2 discussions.}

\begin{table}[htbp]
\centering
\renewcommand{\arraystretch}{1.25}
\caption{ {Derived present-day equation-of-state value $w_0$ and effective CPL-like slope $w_a^{\rm eff}$ for the full $H(z)$+DESI DR2+DES-Dovekie+RSD data combination. The effective slope is obtained by fitting $w_{\rm de}(a)$ to $w_0+w_a(1-a)$ over $0\le z\le2.5$ with $w_0$ fixed for each posterior sample.}}
\label{tab:w0_wa_eff}
\begin{tabular}{ccc}
\hline
\hline
 {Model} &  {$w_0$} &  {$w_a^{\rm eff}$}\\
\hline
 {HDE} &  {$-0.898^{+0.038}_{-0.048}$} &  {$0.398^{+0.057}_{-0.039}$} \\
 {ADE} &  {$-0.797^{+0.003}_{-0.002}$} &  {$0.168^{+0.001}_{-0.002}$} \\
 {RDE} &  {$-0.807^{+0.031}_{-0.032}$} &  {$0.707^{+0.059}_{-0.049}$} \\
\hline
\hline
\end{tabular}
\end{table}

 {The derived values in Table~\ref{tab:w0_wa_eff} show that all three models project onto quintessence-like present-day values, $w_0>-1$, for the full data combination. The positive $w_a^{\rm eff}$ values indicate that $w_{\rm de}$ becomes less negative toward higher redshift over the plotted interval. HDE gives a near-de Sitter present value with moderate evolution, ADE gives the most restricted and slowly varying evolution, while RDE gives a larger effective slope because its dark-energy density is tied directly to the Ricci-curvature terms in the background expansion. These results should not be compared one-to-one with the posterior parameters of a free CPL fit. In CPL analyses, $w_0$ and $w_a$ are independent phenomenological parameters directly constrained by the data, whereas in HDE, ADE, and RDE the redshift evolution of $w_{\rm de}$ is fixed once the model parameters are specified. Therefore, $w_a^{\rm eff}$ is only a low-redshift projection of each model trajectory onto the CPL form, not a replacement for a full CPL analysis.}

 {This comparison is nevertheless useful because it shows whether the physically or geometrically motivated model trajectories overlap qualitatively with the dynamical-dark-energy behavior usually discussed in DESI DR2 CPL analyses \cite{2025PhRvD.112h3515A}. The full-data results in Table~\ref{tab:w0_wa_eff} give $w_0>-1$ and $w_a^{\rm eff}>0$ for all three models, corresponding to restricted quintessence-like trajectories over $0\le z\le2.5$. Thus, these holographic-inspired models do not reproduce the full freedom of CPL-like evolution. Instead, they test whether more constrained physical or geometrical mechanisms, based on a future event horizon, the conformal age of the Universe, or the Ricci scalar, can remain viable under the same late-time data. Since we do not perform a full CPL fit with the same five data combinations, we do not claim that these holographic-inspired models are statistically preferred over CPL.}

\section{Conclusions}

We have performed a comparative late-time constraint analysis of three holographic-inspired dark energy models, namely HDE, ADE, and RDE. The analysis combines cosmic chronometer $H(z)$ measurements, SNe Ia samples, DESI DR2 BAO data, and  {an extended RSD growth compilation}. Five data combinations are considered: $H(z)+$Pantheon+, $H(z)+$DESI DR2+Pantheon+, $H(z)+$DESI DR2+DES-Dovekie, $H(z)+$DESI DR2+DESY5, and  {$H(z)+$DESI DR2+DES-Dovekie+RSD}.  {The sound horizon $r_d$ is treated as a free parameter in the BAO-included combinations, so the constraints are based on late-time information without imposing a CMB-calibrated sound-horizon prior.}

For HDE, the inclusion of DESI DR2 BAO drives the holographic parameter toward $c\simeq1$, indicating an expansion history close to the de Sitter boundary in the future dark-energy-dominated limit.  {Physically, this means that the event-horizon infrared cutoff selected by the data lies near the boundary between phantom-like and quintessence-like HDE evolution.} The constrained Hubble constant remains around $H_0\simeq67$--$68~\Mpc$, close to the Planck $\LCDM$ value and well below the SH0ES measurement. Therefore, HDE does not significantly alleviate the Hubble tension in the present late-time analysis. At the same time, HDE remains statistically competitive with $\LCDM$, although the BIC values show that the evidence is not decisive after accounting for the additional parameter.

For ADE, the agegraphic parameter is consistently constrained around $n\simeq2.8$ once DESI DR2 BAO is included, in agreement with standard NADE benchmark results.  {This stable value reflects the conformal-age scale required by the late-time distance data, while the derived rather than independently sampled $\Omega_{m0}$ makes ADE more predictive but less flexible.} However, ADE prefers a relatively low Hubble constant, $H_0\simeq66.5$--$66.8~\Mpc$, and is generally less competitive than $\LCDM$ according to the information criteria.

For RDE, the Ricci parameter is tightly constrained around $\gamma\simeq0.53$--$0.55$, and the model can improve the best-fit quality according to AIC. However, this improvement comes with a persistent preference for a low matter density, $\Omega_{m0}\simeq0.215$--$0.219$, substantially below the Planck $\LCDM$ value.  {This indicates that the curvature-linked dark-energy component can mimic part of the late-time distance behavior that would otherwise be attributed to the matter component, producing a good background fit at the price of a low matter fraction.}  {The RSD-included result provides a growth-level check and yields $\sigma_{8,0}\simeq0.96$ for RDE, but it does not prove perturbative viability. The combination of low $\Omega_{m0}$ and high fitted $\sigma_{8,0}$ should therefore be regarded as a potential tension suggested by previous studies and by the present background-plus-growth consistency test, rather than as a demonstrated incompatibility with structure formation.}

 {We also reconstructed the redshift evolution of $w_{\rm de}(z)$ and found no robust evidence for crossing the phantom divide over the observed redshift range.} HDE approaches a near-de Sitter behavior, ADE remains quintessence-like by construction, and RDE stays above $w_{\rm de}=-1$ for the late-time best-fit region.  {The $H_0$--$r_d$ posterior contours show strong negative correlations in all BAO-included combinations, emphasizing that the inferred $H_0$ values are late-time constraints marginalized over the sound horizon rather than CMB-calibrated determinations. Since the fitted $r_d$ values remain close to the standard sound-horizon scale, the models do not reduce the Hubble tension by substantially modifying the BAO ruler.} Future analyses should combine DESI BAO, SNe Ia, cosmic chronometers,  {RSD, weak lensing, full-shape clustering, and CMB likelihoods} to test these models at both background and perturbation levels. Overall, HDE gives the most balanced phenomenological performance among the three holographic-inspired models studied here, although current data do not provide decisive evidence that any of them is superior to $\LCDM$ or resolves the Hubble tension.

\section*{Acknowledgments}
This work was supported by National Key R$\&$D Program of China (No. 2024YFC2207400), and (Grant No. 2022YFC2204602).
\bibliography{Bib}
\end{document}